\documentclass[aps, pra, twocolumn, amsfonts, amssymb]{revtex4}

\usepackage{epsfig}

\begin{document}

\title{Output state in multiple entanglement swapping}

\author{Aditi Sen(De), Ujjwal Sen, and Marek \.Zukowski}
 \affiliation{Instytut Fizyki Teoretycznej i Astrofizyki, Uniwersytet
 Gda\'nski, PL-80-952 Gda\'nsk, Poland}

\begin{abstract}
The technique of quantum repeaters is a promising candidate for sending quantum states over long distances
through a lossy channel. The usual discussions of this technique deals with only a 
finite dimensional Hilbert space. However the qubits with which one  implements
this procedure will ``ride''  on  continuous degrees of freedom of the carrier particles. 
Here we analyze the action of quantum repeaters using a model based on pulsed parametric 
down conversion entanglement swapping. 
Our model contains 
some basic traits of a real experiment. We show that the state created, 
after the use of \emph{any} number of parametric down converters in a series 
of entanglement swappings, 
is 
always an entangled (actually distillable) state, although of a different form
than the one that is usually assumed. Furthermore, the output state always 
violates a Bell inequality. 
\end{abstract}


\maketitle

\def\com#1{{\tt [\hskip.5cm #1 \hskip.5cm ]}}

\section{Introduction}
\label{intro}
Entanglement
cannot be created by local operations and classical communication between the parties. 
However
it was shown in Ref.
\cite{ZZHE} (see also \cite{SougatoSwap}) that there exists an operational
scheme, such that  particles can get entangled without ever having interacted
in the past. One of the intruiging things about this phenomenon, which has been 
called entanglement swapping, is that it shows that one cannot always tell whether particles 
are entangled by looking at their 
``common history''. Or the concept of ``common history'' must be suitably enlarged. 
Note  that 
entanglement swapping process is a specific case of quantum teleportation \cite{tele}.
The first experimental realization of entanglement swapping was reported in 
\cite{Pan}. The experiment was a direct realization of the experimental
procedure given in \cite{ZZHE}, and modified in \cite{ZukowskiSwapProc}.

 As a simple illustration
of this phenomenon, 
consider the situation in which 
Alice and Bob, share the singlet
\(\left|\psi^-\right\rangle = \frac{1}{\sqrt{2}}(\left|01\right\rangle - \left|10\right\rangle)\),
and Alice shares another singlet with Claire. Alice now makes a projection measurement on 
her parts of the two singlets in the Bell basis, given by the states
\begin{equation}
\label{phipm}
\begin{array}{lcl}
\left|\phi^\pm \right\rangle = \frac{1}{\sqrt{2}}(\left|00\right\rangle \pm \left|11 \right\rangle), \\
\left|\psi^\pm\right\rangle = \frac{1}{\sqrt{2}}(\left|01\right\rangle \pm \left|10\right\rangle).
\end{array}
\end{equation}
It is easy to check that if 
 Alice now communicates (over a classical channel) the result of her measurement to Bob and Claire, they will know 
that they share one of the Bell states given by eq. (\ref{phipm}). Note that in principle
 the particles 
of Bob and Claire may not have interacted in the past, although they share entanglement after Alice's classical 
communication to them.

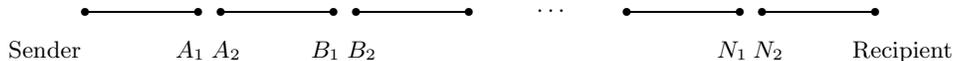
\begin{figure*}[ht]
\begin{center}
\unitlength=0.6mm
\begin{picture}(80,40)(0,0)
\put(-50,28){\line(1,0){25}}
\put(-50,28){\circle*{2}}
\put(-67,18){Sender}
\put(-30,18){\(A_{1}\)}
\put(-22,18){\(A_2\)}
\put(0,18){\(B_1\)}
\put(8,18){\(B_2\)}

\put(90,18){\(N_1\)}
\put(98,18){\(N_2\)}

\put(120,18){Recipient}
\put(50,28){\(\ldots\)}
\put(70,28){\circle*{2}}
\put(70,28){\line(1,0){25}}
\put(95,28){\circle*{2}}
\put(100,28){\circle*{2}}
\put(100,28){\line(1,0){25}}
\put(125,28){\circle*{2}}

\put(-25,28){\circle*{2}}
\put(-20,28){\circle*{2}}
\put(-20,28){\line(1,0){25}}
\put(5,28){\circle*{2}}
\put(10,28){\line(1,0){25}}
\put(10,28){\circle*{2}}
\put(35,28){\circle*{2}}
   
\end{picture}
\end{center}
\caption{Quantum repeaters: A chain of maximally entangled   
states obtained after distilations, are shared between the sender,  
the nodes at \(A\), \(B\), \(\ldots\), \(N\) and the recipient.
Bell measurements are performed at the nodes. This results in an entangled state shared by 
the sender and the reciepent.}
\label{repeatersdhoppic}
\end{figure*}

Apart from this fundamental perspective, entanglement swapping is also important in 
quantum communication   applications.  When sending a quantum state over a noisy channel, the probability 
that it reaches the recipient, decreases with the length of the channel. However, the detectors at the recipient's end
are usually (rather invariably) noisy, and this noise is independent of the length of the channel. Thus after a critical
length, the signal is useless. To circumvent this problem, a proposal was provided \cite{Briegel} that places a number 
of nodes in between the sender and the recipient of the signal. Entangled states are first shared 
in these shorter segments (i.e. between all successive nodes) and thereafter distilled \cite{huge} to obtain 
highly entangled states between all successive nodes (Fig.  \ref{repeatersdhoppic}). 
Finally entanglement swapping is carried out at all nodes 
to obtain highly entangled states between the sender and the recipient
(henceforth called Alice and Bob respectively). It was shown \cite{Briegel} that 
this procedure, called quantum repeaters,
 would lead to highly entangled states between the ends of a noisy channel  of \emph{arbitrary} length, with only 
a polynomial increase in time and logarithmic increase in local resources.

Usual discussions on quantum repeaters deal with only a finite dimensional Hilbert space.
But the qubits with which one  implements
this procedure will ``ride''  on  continuous degrees of freedom of the carrier particles. 
In this paper, we address the problem 
of implementation of this procedure of quantum repeaters, for
the entangled states  prepared between 
successive nodes by spontaneous pulsed parametric down conversion
 \cite{ZZHE,ZukowskiSwapProc}. The
discussion of the process will follow the ones of Refs.
\cite{ZZHE, ZukowskiSwapProc}. The experimental realization of
entanglement swapping fully confirmed the validity of this description \cite{Pan}.
 Actually
in the experiment, polarization entanglement was utilized. But
it is elementary to show the equivalence of such experiment 
with ones involving path entanglement, which will be
our model here (see Ref. \cite{ZHWZ}).   In this paper, 
we assume  that the noise in the channel from a parametric 
down conversion crystal to the nearest nodes is negligible. 
Entanglement swapping is  
carried out at all the nodes.

The description of entanglement swapping that we consider in this paper, 
is still a toy model. Nevertheless it contains 
some basic traits of a possible real experiment. 
We show that the final state prepared between Alice and Bob is 
a so-called maximally correlated state, which is always entangled (actually distillable)
 and always violates a Bell inequality. 
For a wide range of pulses and filters, 
including Gaussian pulses and filters, the output two qubit 
state created between Alice and Bob turns out to be a mixture of two Bell states,
\[\frac{1+V}{2}\left|\phi^+\right\rangle \left\langle \phi^+ \right|
+ \frac{1-V}{2}\left|\phi^-\right\rangle \left\langle \phi^- \right|, 
\]
where \(\left|\phi^\pm \right\rangle\) are given by eq. (\ref{phipm})
and where 
 \(0 \leq V \leq 1\).

\section{Double entanglement swapping}
\label{repeaters}
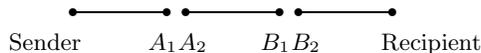
\begin{figure}[ht] 
\begin{center}
\unitlength=0.5mm
\begin{picture}(120,25)(0,0)
\put(0,20){\line(1,0){25}}
\put(0,20){\circle*{2}}
\put(-17,10){Sender}

\put(20,10){\(A_1\)}
\put(28,10){\(A_2\)}
\put(50,10){\(B_1\)}
\put(58,10){\(B_2\)}
\put(82,10){Recipient}
\put(25,20){\circle*{2}}
\put(30,20){\circle*{2}}
\put(30,20){\line(1,0){25}}
\put(55,20){\circle*{2}}
\put(60,20){\line(1,0){25}}
\put(60,20){\circle*{2}}
\put(85,20){\circle*{2}}
\end{picture}
\end{center}
\caption{Quantum repeaters: A chain of three states.}
\label{threeswap}
\end{figure}

Consider the case of three parametric down conversion crystals producing three 
entangled states between 
the sender (Alice) and \(A_1\), \(A_2\) and \(B_1\), and \(B_2\) and 
the recipient (Bob) (see Fig. \ref{threeswap})\cite{footnote_noi}.

\begin{figure*}[ht]
\begin{center}
\unitlength=0.4mm
\begin{picture}(80,60)(0,0)
\put(-95,-19){\line(1,1){10}}
\put(-102,-23){\(\phi^{'}\)}
\put(-100,-5){\vector(-1,-1){10}}
\put(-100,-5){\vector(-1,1){10}}
\put(-105,-4.5){\line(1,0){10}}
\put(-100,-5){\line(1,1){20}}
\put(-100,-5){\line(1,-1){20}}
\put(-85,16){\line(1,0){10}}
\put(-85,-25){\line(1,0){10}}
\put(-60,-5){\line(-1,-1){20}}
\put(-60,-5){\line(-1,1){20}}
\put(-60,0){\line(1,0){20}}
\put(-61,1){\(\scriptscriptstyle{PDCIII}\)}
\put(-60,0){\line(0,-1){10}}
\put(-40,0){\line(0,-1){10}}
\put(-60,-10){\line(1,0){20}}
\put(-40,-5){\vector(1,1){45}}
\put(-40,-5){\vector(1,-1){45}}
\put(-5,25){\line(0,1){10}}
\put(-5,-45){\line(0,1){10}}
\put(30,-5){\vector(-1,-1){45}}
\put(30,-5){\vector(-1,1){45}}
\put(30,0){\line(1,0){20}}
\put(31,1){\(\scriptscriptstyle{PDCII}\)}
\put(30,0){\line(0,-1){10}}
\put(50,0){\line(0,-1){10}}
\put(30,-10){\line(1,0){20}}
\put(50,-5){\vector(1,1){45}}
\put(50,-5){\vector(1,-1){45}}
\put(85,25){\line(0,1){10}}
\put(85,-45){\line(0,1){10}}
\put(120,-5){\vector(-1,-1){45}}
\put(120,-5){\vector(-1,1){45}}
\put(120,0){\line(1,0){20}}
\put(122,1){\(\scriptscriptstyle{PDCI}\)}
\put(120,0){\line(0,-1){10}}
\put(140,0){\line(0,-1){10}}
\put(120,-10){\line(1,0){20}}
\put(140,-5){\line(1,1){20}}
\put(140,-5){\line(1,-1){20}}
\put(155,16){\line(1,0){10}}
\put(155,-25){\line(1,0){10}}
\put(180,-5){\line(-1,-1){20}}
\put(180,-5){\line(-1,1){20}}
\put(175,-4.5){\line(1,0){10}}
\put(180,-5){\vector(1,1){10}}
\put(180,-5){\vector(1,-1){10}}
\put(165,-10){\line(1,-1){10}}
\put(175,-25){\(\phi\)}
\put(195,-20){\(a_{+}\)}
\put(195,5){\(a_{-}\)}
\put(-125,-20){\(f_{+}\)}
\put(-125,5){\(f_{-}\)}
\put(188, -5){\(\scriptscriptstyle{BS}\)}

\put(-93, -5){\(\scriptscriptstyle{BS}\)}

\put(-82, 18){\(\scriptscriptstyle{M}\)}

\put(157, 18){\(\scriptscriptstyle{M}\)}

\put(-82, -30){\(\scriptscriptstyle{M}\)}

\put(157, -30){\(\scriptscriptstyle{M}\)}

\put(-70,-20){\(f\)}
\put(-70,5){\(f^{'}\)}
\put(-40,-24){\(e^{'}\)}
\put(-35,5){\(e\)}

\put(23,-20){\(d^{'}\)}
\put(20,5){\(d\)}

\put(55,-20){\(c\)}
\put(55,5){\(c^{'}\)}

\put(110,-20){\(b\)}
\put(110,5){\(b^{'}\)}

\put(140,-24){\(a^{'}\)}
\put(145,5){\(a\)}

\put(0,-40){\(\scriptscriptstyle{BS}\)}

\put(0, 30){\(\scriptscriptstyle{BS}\)}

\put(90,30){\(\scriptscriptstyle{BS}\)}

\put(90,-40){\(\scriptscriptstyle{BS}\)}

\put(100,35){\(i_1\)}
\put(110,35){\({\star}\)}
\put(65,35){\(i_1^{'}\)}
\put(100,-55){\(i_2\)}
\put(110,-55){\({\star}\)}
\put(65,-55){\(i_2^{'}\)}
\put(7,-55){\(i_4\)}
\put(17,-55){\({\star}\)}
\put(7,35){\(i_3\)}
\put(17,35){\({\star}\)}
\put(-23,-55){\(i_4^{'}\)}
\put(-23,35){\(i_3^{'}\)}
 
\end{picture}
\end{center}
\vspace{1.8cm}
\caption{Entanglement swapping in a chain of three parametric down converters. The boxes denoted
as PDC-I, PDC-II, and PDC-III symbolizes the full parametric down conversion device, including the pump laser,
crystal, a set of apertures that select suitable output directions, as well as suitably oriented
mirrors that direct the radiation into the interferometers shown in the figure. 
The radiation at the detectors 
\(i_1, \ldots, i_4\) and \(i_1^{'}, \ldots, i_4^{'}\) is observed by detectors. The interferometers are
built out of mirrors \(M\) and 50-50 beamsplitters (BS). We need to
 put suitable frequency filters 
in front of all BS leading   to 
the internal outputs \(i_{(.)}\) and \(i_{(.)}^{'}\). 
The symbols \(\phi\) and \(\phi^{'}\) denote the local phase shifters. For simplicity, 
we shall study the case when only  four internal detectors (the ``i''s)
 fire (those marked with asterix).
Due to the specific properties of downconversion 
\cite{ZZHE, ZukowskiSwapProc}, no filters are needed in the output beams of the device, 
\(a\) and \(f\).} 
\label{PDCpict}
\end{figure*}
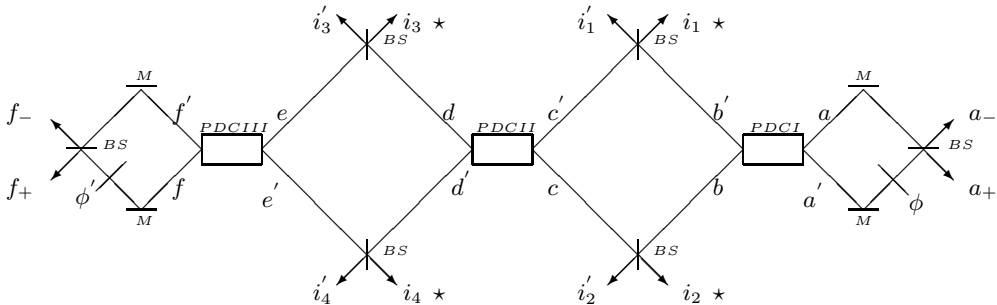

Entanglement swapping will be carried out at \(A_1 A_2\) and \(B_1 B_2\). 
In Fig. \ref{PDCpict}, we  give a 
schematic description of entanglement swapping
in an array of three type I spontaneous parametric down conversions (PDC) \cite{ZZHE,ZukowskiSwapProc}.
Note that frequency filters are in front of every beamsplitter, in the ``internal'' part of the device. They are necessary to
make the photons emerging independently from two different sources, indistinguishable
by the detectors behind the beamsplitters (for the physical reasons for this, 
see \cite{ZZHE, ZukowskiSwapProc}).

We make the \emph{simplifying assumptions}
that the optical lengths of all source-detector paths are equal and that phase shifters work in the range
 of the order of the wavelength (i.e. between 0 and \(2 \pi\)).
This enables us to neglect all retardation effects.

The description of the two-photon initial entangled state 
will depend 
on whether 
the corresponding PDC is an ``external'' or an ``internal'' one, in the series of PDCs. In any series 
of PDCs, there will be two external ones, while the rest will be called internal. For example, in the case of
three PDCs, as described in Fig. \ref{PDCpict}, PDC-I and PDC-III are externals, 
while PDC-II is an internal one. 
If the ``idler'' photon emitted by the external PDCs  in Fig. \ref{PDCpict} (produced by a single  pulse
from a laser pump), manages to pass via the filters, the resulting
two-photon state is given by (see Appendix)
\begin{widetext}
\begin{equation}
\label{pdcstate1}
\left| \psi_{ext} (x,y,x^{'},y^{'}) \right\rangle=  \int d\omega_i \int d \omega_s \int d \omega_p F(\omega_p) 
          \Delta (\omega_p - \omega_i - \omega_s)
           f(\omega_i) 
           (a_x^{\dagger}(\omega_i)a_y^{\dagger}(\omega_s) + 
            a_{x^{'}}^{\dagger}(\omega_i)a_{y^{'}}^{\dagger}(\omega_s) ) \left|\Omega \right\rangle,  
\end{equation}
\end{widetext}
where \(a_x^{\dagger}(\omega_i)\) (\(a_y^{\dagger}(\omega_s)\))
and \(a_{x^{'}}^{\dagger}(\omega_i)\) (\(a_{y^{'}}^{\dagger}(\omega_s)\))  
are the  creation operators of photons  of the idler (signal)
of frequency \(\omega_i\) (\(\omega_s\)) respectively in beams 
\(x\) and \(x^{'}\) (\(y\) and \(y^{'}\)). The function \(F\)
represents the spectral content of the pump pulse of frequency \(\omega_p\) and 
\(f\) is the transmission function of the filters and is assumed to be centered at 
\( \omega_0/2\), where in turn, \(\omega_0\) is the central frequency of the laser
pump pulse. The function 
\(\Delta(\omega_p - \omega_i - \omega_s)\) is due to the phase-matching condition
 of the PDC process. \(\left|\Omega \right\rangle\) is the vacuum state.
The state produced at PDC-I 
is \(\left| \psi_{ext} (b,a,b^{'},a^{'}) \right\rangle\) and that at PDC-III is 
\(\left| \psi_{ext} (f,e,f^{'},e^{'}) \right\rangle\).
We assume, in all our considerations,  the perfect case and so we will replace our \(\Delta\) 
by the  
Dirac delta function.  
Here and henceforth,
unless stated otherwise, we ignore normalization of states.

If the two photons  produced  by  the internal PDC (PDC-II) in Fig. \ref{PDCpict},
 manage to pass the filters, their state
acquires the following form  (see Appendix):
\begin{widetext}
\begin{equation}
\label{pdcstate2}
\left| \psi_{int}(d,c,d^{'},c^{'}) \right\rangle =  \int d\omega_i^{'} \int d \omega^{'}_s \int d \omega^{'}_p F(\omega^{'}_p) 
          \Delta (\omega^{'}_p - \omega^{'}_i - \omega^{'}_s)
           f(\omega^{'}_i) f(\omega^{'}_s) 
           (a_d^{\dagger}(\omega_i^{'})a_c^{\dagger}(\omega_s^{'}) +  
            a_{d^{'}}^{\dagger}(\omega_i^{'})a_{c^{'}}^{\dagger}(\omega_s^{'}) ) \left|\Omega \right\rangle.  
\end{equation}
\end{widetext}
Note that there is an extra filter function in this case. This is a signature of 
the fact that both the photons 
from PDC-II are  detected by the \emph{internal} detectors  and to reach there, they
must pass the filters. 
In the general case of an arbitrary number (say, \(n\)) of PDCs in a series, 
there will be two entangled pairs 
described by  eq. (\ref{pdcstate1}), while there will be \(n-2\) ``internal'' entangled pairs 
described by eq. (\ref{pdcstate2}). Note here that previous works on entanglement 
swapping with \emph{two} PDCs \cite{ZZHE, ZukowskiSwapProc} did not need to consider 
any ``internal'' entangled pairs.

In the case of entanglement swapping by \( 3\) PDCs (as shown in Fig. \ref{PDCpict}), 
our initial state, if all photons manage to pass the filters,  
is given by 
\[\begin{array}{lcl}
&&\left|\Psi \right \rangle = \\
 && \left |\psi_{ext} (f,e,f^{'},e^{'})\right \rangle
           \left |\psi_{int}(d,c,d^{'},c^{'}) \right\rangle
                \left |\psi_{ext} (b,a,b^{'},a^{'}) \right\rangle.
\end{array}
\]
 Now suppose that the  detector \(i_1\) fires at time \(t_1\), \(i_2\) at time \(t_2\),
\(i_3\) at time \(t_3\), and \(i_4\) at time \(t_4\). Then  the wave packet collapses 
into the  state 
\begin{equation}
\label{idlerstate}
\left|\Psi; t_1, t_2, t_3, t_4 \right \rangle =
i_1(t_1)  i_2(t_2)  i_3(t_3)  i_4(t_4)  \left|\Psi \right \rangle
\end{equation}
where, for example, 
\[
i_1(t_1) = \int d \omega \exp \{-i \omega t_1\} a_{i_1}(\omega).
\]
Note that the scalar product of 
\(\int d \omega \exp \{i \omega t_1\} a_{i_1}^{\dagger}(\omega)\)  with a single photon
state \[\int d\omega g(\omega) a^{\dagger}_{i_1}(\omega) \left|\Omega\right\rangle \]
gives the probability amplitude to detect this photons at time \(t_1\).
Let us assume that our 50-50 beamsplitters (BS) are the symmetric ones.
That is, one has, e.g. 
\[a_{i_{1}}^{\dagger}(\omega) = \frac{1}{\sqrt{2}}\left( a_{c^{'}}^{\dagger}(\omega) + 
                                i a_{b^{'}}^{\dagger}(\omega)\right).\]
Note that the creation operator of the reflected beam always enters the relation for \(a_{i_k}^{\dagger}\)
with an \(i\) factor.
As we consider here the idealized case, we substitute  \(\Delta\) by  Dirac delta functions to obtain
\begin{widetext}
\begin{equation}
\label{idlerstatesimp}
\left| \Psi; t_1, t_2, t_3, t_4 \right\rangle =
\int dt F^*(t) f(t-t_4)f(t-t_1) a_f^{\dagger}[g,t_3]a_a^{\dagger}[g,t_2] 
+ \int dt F^*(t) f(t-t_3)f(t-t_2) a_{f^{'}}^{\dagger}[g,t_4]a_{a^{'}}^{\dagger}[g,t_1],
\end{equation}
\end{widetext}
where we denote for example,
\[
a_f^{\dagger}[g,t_3] = \int d \omega_s g(\omega_s,t_3) a_f^{\dagger}(\omega_s)
\]
with
\[
g(\omega_s,t_3) = \int dt \exp \{i \omega_s t\}F^*(t)f(t-t_3).
\]

After the photons in the state \(\left| \Psi; t_1, t_2, t_3, t_4 \right\rangle\) 
pass through the phase shifters (as depicted in Fig. \ref{PDCpict}), the new state, 
\(\left| \Psi; t_1, t_2, t_3, t_4, \phi, \phi^{'} \right\rangle\),  is obtained 
by replacing \(a^{\dagger}_{a^{'}}\) by \(\exp \{i \phi\} a^{\dagger}_{a^{'}}\)
and \(a^{\dagger}_{f}\) by \(\exp \{i \phi^{'}\} a^{\dagger}_{f}\).
Let us assume now that the photons (after passing 
through phase shifters and 50-50 beam splitters (cf. Fig. \ref{PDCpict}))
are detected at   \(a_{+}\) and \(f_{+}\)  at times 
\(t_a\) and \(t_f\) respectively. Hence the amplitude of such a process is
\[\begin{array}{lcl}
 &&A_{a_+,f_+}(t_a, t_f, t_1, t_2, t_3, t_4) =\\
 &&\left\langle f_+(t_f) , a_+(t_a) \right|
\left| \Psi; t_1, t_2, t_3, t_4, \phi, \phi^{'} \right\rangle,
\end{array}
\]
where, for example,
\[\begin{array}{lcl}
\left|f_+(t_f), a_+(t_a)  \right\rangle &=& 
\int d \omega \exp \{i \omega t_f\} a^{\dagger}_{f_+}(\omega)\\
&& \int d \omega^{'} \exp \{i \omega^{'} t_a\} a^{\dagger}_{a_+}(\omega^{'})
\left|\Omega\right\rangle .
\end{array}\]
Writing it out explicitly, one obtains 
\[ A_{a_+, f_+} = \exp\{i(\phi - \phi^{'})\} T_1  + T_2,\] 
where 
\begin{equation}
\begin{array}{lcl}
\label{amplitude3}
&& T_1 = T(t_a, t_f, t_1, t_2, t_3, t_4)  \\
&& \equiv \int dt dt^{'} dt^{''} 
            \int d \omega_s  d\omega_s^{''} \exp\{i(\omega^{''}_s t + \omega_s t^{'})\}\times \\
&&                      \exp\{-i(\omega^{''}_s t_{f} + \omega_s t_{a})\} 
                 F^*(t) F^*(t^{'})F^*(t^{''}) \times  \\
    &&             f(t-t_3)  f(t^{'}- t_2) f(t^{''}- t_4) f(t^{''}- t_1)
\end{array}
\end{equation}
and \[T_2 = T(t_a, t_f, t_2, t_1, t_4, t_3).\]
However, due to the finite time resolution of the detectors,
the precise detection times are not known. Therefore the full probability, \(P\), 
of the process is obtained
after integration of the square of the amplitude over the detection times:
\begin{equation}
\label{prob}
\begin{array}{lcl}
&&P(a_+, f_+) =  \int [|T_1|^2 + |T_2|^2  \\
&& + 2 \cos(\phi - \phi^{'} + \phi_{0}) |T_1  T_2^* |]  dt_1 dt_2 dt_3 dt_4 dt_f dt_a
\end{array}
\end{equation}
(where \(\phi_{0}\) is a phase). We integrate here 
from \(- \infty \) to \(+\infty\), because the time resolution of detectors 
is usually by orders of magnitude bigger than the duration of the pumping pulse.
We can now calculate the interferometric contrast, or the visibility 
\[
V_3 = \frac{\max P - \min P}{\max P + \min P},
\] 
of the two particle interference process as observed by changing the values of the local phase shifts in the 
two external interferometers. 
(The index \(3\) stands for the number of
PDCs involved.)
The resulting value is
given by
\begin{equation}
\label{vis3}
V_3 = \frac{2 \int |T_1  T_2^* |  dt_1 dt_2 dt_3 dt_4 dt_f dt_a}{\int(|T_1|^2 + |T_2|^2) dt_1 dt_2 dt_3 dt_4 dt_f dt_a}.
\end{equation}

 We will now write down the explicit forms of the expressions in the numerator and denominator of \(V_3\).
We have
\begin{equation}
\begin{array}{lcl}
\label{T_1T_2^*3next}
F_{1234} &\equiv & \int |T_1T_2^{*}| dt_1 dt_2 dt_3 dt_4 dt_f dt_a \\
&      =  &  \int dt_1 dt_2 dt_3 dt_4 |\int dt dt ^{'} dt^{''} d\overline{t^{''}} \times \\
                &&   |F(t)|^2 |F(t^{'})|^2 F^*(t^{''}) F(\overline{t^{''}}) \times\\
                &&    f(t-t_3) f^{*}(t - t_4)
                      f(t^{'}- t_2) f^{*}(t^{'} - t_1)  \times\\
                   &&  f(t^{''}- t_4) f(t^{''}- t_1)  f^{*}(\overline{t^{''}}- t_3) f^{*}(\overline{t^{''}}- t_2)|.
 \end{array}
\end{equation}
Throughout the paper, a ``bar'' will represent a new variable. This should not be confused with complex conjugation,
which we  denote here by a ``\(^*\)''.

A similar simplifying gives
\begin{equation}
\begin{array}{lcl}
\label{T_1T_1^*3}
 G_{1234} &\equiv & \int |T_1|^2 dt_1 dt_2 dt_3 dt_4 dt_f dt_a \\
  &= & \int dt dt ^{'} dt^{''} d\overline{t^{''}} \int dt_1 dt_2 dt_3 dt_4 \times \\
           &&         |F(t)|^2 |F(t^{'})|^2 F^*(t^{''}) F(\overline{t^{''}}) \times \\
             &&    |f(t-t_3)|^2 |f(t^{'}- t_2)|^2
              f(t^{''}- t_4) f(t^{''}- t_1)  \times \\
&&      f^{*}(\overline{t^{''}}- t_4) f^{*}(\overline{t^{''}}- t_1).
 \end{array}
\end{equation}
Note that in \(G_{1234}\), the indices are  ordered. 
One can easily find that  
\begin{equation}
\label{T_2T_2^*3}
\int |T_2|^2 dt_1 dt_2 dt_3 dt_4 dt_f dt_a    =    G_{2143}.
\end{equation}

Hence the visibility of the two particle interference that can be obtained due to the two-fold entanglement
swapping (Fig. \ref{PDCpict}), is given by 
\begin{equation}
\begin{array}{rcl}
\label{vis3G}
 V_3= \frac{2 {F_{1234}}}
             {G_{1234} + G_{2143}}.
 \end{array}
\end{equation}

\section{Arbitrary number of swappings}
\label{arbitrary}

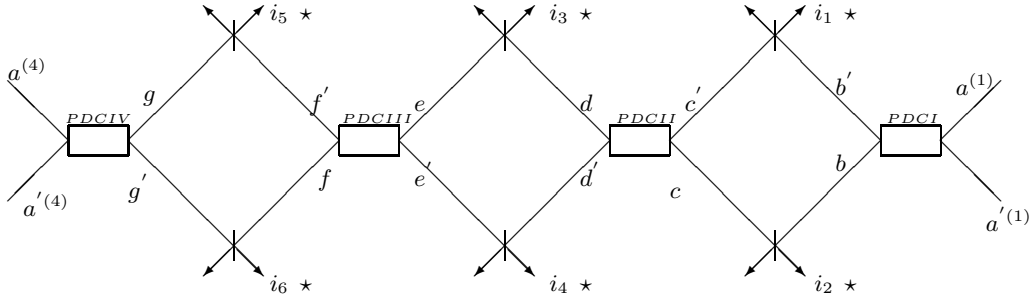
\begin{figure*}[ht]
\begin{center}
\unitlength=0.4mm
\begin{picture}(160,60)(0,0)
\put(-60,-5){\line(-1,-1){20}}
\put(-60,-5){\line(-1,1){20}}
\put(-60,0){\line(1,0){20}}
\put(-61,1){\(\scriptscriptstyle{PDCIV}\)}
\put(-60,0){\line(0,-1){10}}
\put(-40,0){\line(0,-1){10}}
\put(-60,-10){\line(1,0){20}}
\put(-40,-5){\vector(1,1){45}}
\put(-40,-5){\vector(1,-1){45}}
\put(-5,25){\line(0,1){10}}
\put(-5,-45){\line(0,1){10}}
\put(30,-5){\vector(-1,-1){45}}
\put(30,-5){\vector(-1,1){45}}
\put(30,0){\line(1,0){20}}
\put(31,1){\(\scriptscriptstyle{PDCIII}\)}
\put(30,0){\line(0,-1){10}}
\put(50,0){\line(0,-1){10}}
\put(30,-10){\line(1,0){20}}
\put(50,-5){\vector(1,1){45}}
\put(50,-5){\vector(1,-1){45}}
\put(85,25){\line(0,1){10}}
\put(85,-45){\line(0,1){10}}
\put(120,-5){\vector(-1,-1){45}}
\put(120,-5){\vector(-1,1){45}}
\put(120,0){\line(1,0){20}}
\put(122,1){\(\scriptscriptstyle{PDCII}\)}
\put(120,0){\line(0,-1){10}}
\put(140,0){\line(0,-1){10}}
\put(120,-10){\line(1,0){20}}
\put(140,-5){\vector(1,1){45}}
\put(140,-5){\vector(1,-1){45}}
\put(175,25){\line(0,1){10}}
\put(175,-45){\line(0,1){10}}
\put(210,-5){\vector(-1,-1){45}}
\put(210,-5){\vector(-1,1){45}}
\put(210,0){\line(1,0){20}}
\put(210,-10){\line(0,1){10}}
\put(210, -10){\line(1,0){20}}
\put(230, -10){\line(0,1){10}}
\put(230, -5){\line(1, 1){20}}
\put(230, -5){\line(1, -1){20}}

\put(212,1){\(\scriptscriptstyle{PDCI}\)}

\put(195,-15){\(b\)}
\put(195,10){\(b^{'}\)}
\put(245, -35){\(a^{'(1)}\)}
\put(235, 10) {\(a^{(1)}\)}
\put(-75,-30){\(a^{'(4)}\)}
\put(-80,15){\(a^{(4)}\)}
\put(-40,-24){\(g^{'}\)}
\put(-35,8){\(g\)}

\put(23,-20){\(f\)}
\put(20,5){\(f^{'}\)}

\put(55,-20){\(e^{'}\)}
\put(55,5){\(e\)}

\put(110,-20){\(d^{'}\)}
\put(110,5){\(d\)}

\put(140,-24){\(c\)}
\put(145,5){\(c^{'}\)}

\put(188, 35){\(i_1\)}
\put(198, 35){\({\star}\)}

\put(188,-55){\(i_2\)}
\put(198, -55){\({\star}\)}
\put(100,35){\(i_3\)}
\put(110,35){\({\star}\)}
\put(100,-55){\(i_4\)}
\put(110,-55){\({\star}\)}
\put(7,-55){\(i_6\)}
\put(17,-55){\({\star}\)}
\put(7,35){\(i_5\)}
\put(17,35){\({\star}\)}
 
\end{picture}
\end{center}
\vspace{1.8cm}
\caption{An exemplary extended chain of swappings. Imagine that the initial states
emitted by the four sources were entangled two photon states, with the property 
that the one photon is emitted to the right and other one to the left, and 
the emissions are such that both photons are in primed beams or both are in the unprimed ones. 
Now if the detectors \(i_1, \ldots, i_6\) fired (the ones with asterix), one has only two
consistent possibilities.
(i)  \(i_1\) fired due to source 1, but then \(i_2\) fired due to source 2, 
therefore \(i_3\)
fired again due to source 2 whereas \(i_4\) fired thanks to source 3, thus
\(i_5\) fired also because of source 3, and finally \(i_6\) fired because of source 4. 
This implies that a photon is in the beam \(a^{(4)}\)  and in the beam \(a^{'(1)}\).
(ii) The second possibility is that the \(i_1\) firing is due to source 2 and etc. 
Thus the other element of the final superposition must consist of photons  in
beams \(a^{'(1)}\)  and in the beam \(a^{(4)}\).} 
\label{4PDCpict}
\end{figure*}

Consider now a chain (i.e., arranged in a series) of \emph{any} number, \(n\),
 of PDCs (see Fig. \ref{repeatersdhoppic}).
Let \(a^{(1)}\) and \(a^{'(1)}\) denote the paths within the external interferometer
into which one of the photons from the first PDC enters. The paths within the other external 
interferometer, into which enters the photon from the last PDcs will 
be denoted as
 \(a^{(n)}\) and \(a^{'(n)}\). 
Let \(\omega_s^{(1)}\) and \(\omega_s^{(n)}\) be the corresponding frequencies
 of the photons. If all the photons manage through the frequency filters, the state is
\begin{equation}
\label{product}
\left|\psi^{(1)}\right\rangle \otimes \left(\otimes_{l=2}^{n-2} 
\left|\psi^{(l)}\right\rangle \right) \otimes \left|\psi^{(n)}\right\rangle
\end{equation}
with the states \(\left|\psi^{(1)}\right\rangle\) and \(\left|\psi^{(n)}\right\rangle\) 
of the type characteristic for the external PDCs and \(\left|\psi^{(l)}\right\rangle\)
(\(l = 2, \ldots, n-2\))
for the internal ones (compare eqs. (\ref{pdcstate1}) and (\ref{pdcstate2})). Imagine now that
again all internal detectors \(i_{k}\), \(k=1, 2, \ldots, 2n-2\) fire. The final state
into which the state in eq. (\ref{product}) collapses is of the following form:
\begin{equation}
\label{any}
\begin{array}{rcl}
\left| \chi_t \right\rangle &=& \int d\omega^{(1)}_s d\omega^{(n)}_s
( X_t(\omega^{(1)}_s, \omega^{(n}_s) 
a_{a^{(n)}}^{\dagger} (\omega^{(n)}_s) a_{a^{'(1)}}^{\dagger} (\omega^{(1)}_s) \\
&+& 
Y_t(\omega^{(1)}_s, \omega^{(n)}_s) 
a_{a^{'(n)}}^{\dagger} (\omega^{(n)}_s) a_{a^{(1)}}^{\dagger} (\omega^{(1)}_s))
\left|\Omega \right\rangle.
\end{array}
\end{equation}
The functions \(X_t\) and \(Y_t\) depend on 
the detection times
at the internal detectors (along with the frequencies). The subscript \(t\) stands for the
full set of  times of detection at the internal detectors, that is 
\(t = t_1, t_2, \ldots, t_{2n-2}\). To see this structure of the state for \(n=4\), 
see Fig. \ref{4PDCpict}. 

The  effective mixed state, \(\rho^{swap}\), that we obtain, is an incoherent sum of the state \(\chi_t\),
with the sum (integration) being 
over the detection times:
\[
\rho^{swap} = \int dt \left| \chi_t \right\rangle \left\langle \chi_t \right|,\]
 where 
\(dt \equiv \Pi_{i=1} ^{2n} dt_i\).

Again we now assume that optical lengths of all source-detector paths are equal and that phase shifts 
\(\phi\) and \(\phi^{'}\) 
are of the order of the wavelength (i.e. between 0 and \(2 \pi\)). Since the two output photons are fully
distinguishable (i.e. their sources are known), it would be no harm to abandon the second
quantized description. Thus, one can rewrite the formula  (\ref{any}) in the following way. Note that
\(a^{\dagger}_{a^{(.)}} (\omega) \left|\Omega \right\rangle\) is a single photon state. The photon is in path \(a^{(.)}\) and 
has frequency \(\omega\). Therefore from the first quantized point of view, this state can be replaced
by the tensor product \(\left|a\right\rangle \left |\omega\right\rangle\), where 
\(\left|a\right\rangle\) describes the path variable
 and \(\left |\omega\right\rangle\) the frequency variable (energy). Using such notation,
 eq. (\ref{any}) can be put as 
\begin{equation}
\begin{array}{rcl}
\left|\chi_t\right\rangle 
&=& \int d\omega^{(1)}_s d\omega^{(n)}_s \times\\
&&( X_t(\omega^{(1)}_s, \omega^{(n}_s) 
{\left|a^{'(1)} \right\rangle}_{1} 
{\left|\omega_s^{(1)} \right\rangle}_{1} {\left|a^{(n)} \right\rangle}_{n} 
{\left|\omega_s^{(n)} \right\rangle}_{n}\\
&+& 
Y_t(\omega^{(1)}_s, \omega^{(n)}_s) , 
{\left|a^{(1)} \right\rangle}_{1} 
{\left|\omega_s^{(1)} \right\rangle}_{1} {\left|a^{'(n)} \right\rangle}_{n} 
{\left|\omega_s^{(n)} \right\rangle}_{n} ).
\end{array}
\end{equation}
The frequency degrees of freedom in \(\rho^{swap}\) can be traced out to obtain the
 (unnormalized) state of the path (i.e. the qubit) degrees   of freedom as
\[\begin{array}{lcl}
\rho^{swap}_{path}= \int  d\omega_s^{(1)}  d\omega_s^{(n)}  dt    \quad \zeta
\end{array}
\]
where 
\[
\begin{array}{lcl}
&&\zeta = \\
&&(|X(\omega^{(1)}_s, \omega^{(n)}_s)|^2
 \left|a^{'(1)}\right\rangle \left\langle a^{'(1)}\right|
\otimes \left|a^{(n)}\right\rangle \left\langle a^{(n)}\right| \\
 && + |Y(\omega^{(1)}_s, \omega^{(n)}_s)|^2 \left|a^{(1)}\right\rangle \left\langle 
a^{(1)}\right|
\otimes \left|a^{'(n)}\right\rangle \left\langle a^{'(n)}\right| \\
&& +  [X(\omega^{(1)}_s, \omega^{(n)}_s)Y^{*}(\omega^{(1)}_s, \omega^{(n)}_s) \times \\
&& \left|a^{'(1)}\right\rangle \left\langle a^{(1)}\right|
\otimes \left|a^{(n)}\right\rangle \left\langle a^{'(n)}\right| + \mbox{h.c.}]).
\end{array}
\]
Here h.c. denotes hermitian conjugate of the term before it within square brackets.

After normalization, one gets 
\begin{equation}
\label{ultimate}
\begin{array}{rcl}
\rho_{path}^{swap}&=& \frac{1}{b+c} (b \left|00\right\rangle\left\langle 00\right| +
             c \left|11\right\rangle\left\langle 11\right|  \\
&+&[a\left|00\right\rangle\left\langle 11\right| + \mbox{h.c.}])
\end{array}
\end{equation}
Here
\(a = 
\int d \omega dt XY^*\),
\(b = 
\int d \omega dt |X|^2\),
\(c = 
\int d \omega dt |Y|^2\), and 
\[  \left|0\right\rangle = \left|a^{(1)}\right\rangle, 
\quad   \left|1\right\rangle = \left|a^{'(1)}\right\rangle,\]
and similarly for \(\left|a^{(n)}\right\rangle\) and \(\left|a^{'(n)}\right\rangle\).
Writing \(a=r_a \exp\{i\theta_a\}\) (\(r_a \geq 0\), \(\theta_a\) real), 
and redefining the state \(\left|1\right\rangle\) of 
the first particle as  \(\exp\{-i\theta_a\}\left|a^{'(1)}\right\rangle\), the normalized state
 \(\rho_{path}^{swap}\) reads
\begin{equation}
\label{final}
\begin{array}{rcl}
\rho_{path}^{swap} &  = & \frac{1}{b+c} (b \left|00\right\rangle\left\langle 00\right| +
             c \left|11\right\rangle\left\langle 11\right| \\ 
&+&[r_a\left|00\right\rangle\left\langle 11\right| + \mbox{h.c.}])
\end{array}
\end{equation}
where \(r_a\), \(b\), and \(c\) are all positive.

One can now  check that 
\begin{equation}
\label{calv}
 V_n = \frac{2r_a}{b+c}
\end{equation}
where \(V_n\) is the visibility of the two-photon interference in the 
external interferometers (like those in Fig. \ref{PDCpict}).

Let us give the values of the parameter in eq. (\ref{calv}) for the case of   Fig. \ref{PDCpict}. One has 
\[
\begin{array}{lcl}
&& r_a = \int d \omega_s  d\omega_s^{''}  \int  dt_1 dt_2 dt_3 dt_4  \int dt dt^{'} dt^{''}
             \int d \overline{t} d \overline{t^{'}} d \overline{t^{''}} \times \\
&& \exp\{i(\omega_s^{''} t + \omega_s t^{'})\} 
              F^*(t) F^*(t^{'})F^*(t^{''}) \times \\
&&      f(t-t_3)  f(t^{'}- t_2) f(t^{''}- t_4) f(t^{''}- t_1) \times \\
&& \exp\{i(\omega_s^{''} \overline{t} + \omega_s \overline{t^{'}})\} 
              F^*(\overline{t}) F^*(\overline{t^{'}})F^*(\overline{t^{''}})\times \\
&&f(\overline{t}-t_4)  f(\overline{t^{'}}- t_1) f(\overline{t^{''}}- t_3) f(\overline{t^{''}}- t_2).
\end{array}
\]
Comparing with  eq. (\ref{T_1T_2^*3next}), one can  verify that 
\[r_a = F_{1234}.\]
Similarly using eqs. (\ref{T_1T_1^*3}) and (\ref{T_2T_2^*3}), one finds that 
\(b\) and \(c\) are respectively \(G_{1234}\) and \(G_{2143}\). Using eq. (\ref{vis3G}), one therefore 
obtains the relation in eq. (\ref{calv}), for the case of 
three parametric down conversions. It is straightforward to see that the same relation holds for an arbitrary
number of PDCs. 
As we have mentioned earlier, the case of \emph{two} PDCs is slightly different,
in the sense that there are no internal PDCs. However it is easy to check that 
the relation in eq. (\ref{calv}) is true even for the case of two PDCs.

Whenever \(r_a \ne 0\), the state \(\rho_{path}^{swap}\) has nonpositive partial transpose  \cite{PT}.
Therefore  the state  is entangled whenever \(r_a \ne 0\) \cite{Peres1}
(see also \cite{HHHPPT}). The entanglement of formation of \(\chi_{path}\) is \cite{huge, Wootters} 
\begin{equation}
\label{EoF}
{\cal E} = H(\frac{1}{2}(1 + \sqrt{1 - V_n^2}))
\end{equation}
where 
\(H(x) = -x\log_2 x - (1-x)\log_2 (1-x)\) is the binary entropy function.
Note that the visibility \(V_n\) is the so-called concurrence \cite{Wootters} of 
the state \(\rho_{path}^{swap}\).
In fact, entanglement of formation is known to be additive for the state 
\(\rho_{path}^{swap}\) \cite{Cirac, ratio}
(see also \cite{Shor}),
and hence the expression displayed in (\ref{EoF}) is 
also the entanglement cost (asymptotic entanglement of formation) of \(\rho_{path}^{swap}\).
States with nonpositive partial transpose in \(2 \otimes 2\)  are 
distillable \cite{HHHdist}.
Therefore the state \(\rho_{path}^{swap}\), being in \(2 \otimes 2\), is also \emph{distillable} whenever 
\(r_a \ne 0\).

Bipartite States of the form 
\[\sum_{ij} a_{ij}\left|ii\right\rangle \left\langle jj \right|
\]
are called maximally correlated states \cite{Rains1}. It is known that for such states in \(2 \otimes 2\),
the entanglement cost 
is strictly higher than its distillable entanglement \cite{Cirac, ratio} (cf. \cite{Rains1, Rains2}). 
We therefore have irreversibility in asymptotic local manipulations of entanglement for such 
states.
The state  \(\rho_{path}^{swap}\) being a  maximally correlated state  in \(2 \otimes 2\),
would also have its entanglement cost  strictly greater than its distillable entanglement.

\begin{figure}[tbp]
  \epsfig{figure=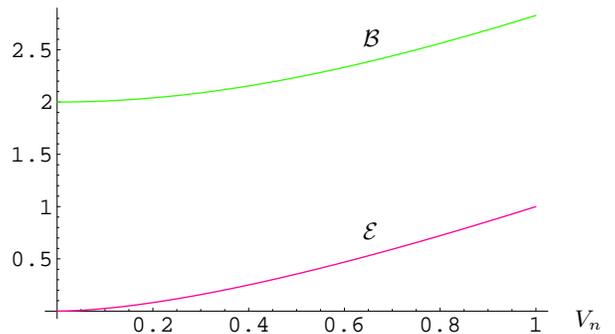,width=0.40\textwidth}
\put(10,5){\(V_n\)} 
\put(-70,38){\({\cal E}\)}
\put(-70,111){\({\cal B}\)}  
\caption{Plot 
of entanglement cost (\({\cal E}\)), given by eq. (\ref{EoF}), and the maximal 
amount of violation of Bell inequality (\({\cal B}\)), given by eq. (\ref{BV}), against 
the visibility \(V_n\) for the state \(\chi_{path}\) obtained after the detection 
of the idlers in the entanglement 
swapping process with \(n\) parametric down conversions.} \label{figur}
\end{figure}

Additionally, the state \(\rho_{path}^{swap}\) 
violates a Bell inequality \cite{Bell}  whenever \(r_a \ne 0\). The maximal amount of violation 
is \cite{HHHBV}
\begin{equation}
\label{BV}
{\cal B} = 2\sqrt{1 + V_n^2}.
\end{equation}  
The \(r_a = 0\) case corresponds to null visibility
and can therefore be ignored. We plot the entanglement of formation (which in our case is entanglement cost)
and the maximal amount of violation of Bell inequality against the visibility  \(V_n\) in Fig. \ref{figur}. 
We therefore obtain the important fact that the output state, resulting from a series of 
entanglement swappings, is always entangled and always violates local realism. 
Earlier works on  entanglement swapping process via \emph{two} parametric down converters
\cite{ZZHE, ZukowskiSwapProc} made the tacit assumption that the output is  a pure state admixed with 
white noise, and reached the conclusion that the  swapped state is separable and does not violate 
local realism for low visibilities.

We will now discuss an interesting point 
with respect to violation of Bell inequalities of the state \(\rho^{swap}_{path}\).
Let us, for this state, find 
the \(T\)-matrix, whose elements for any state \(\varrho\), of two qubits, is given by 
\[T_{ij} = \mbox{tr}(\sigma_i \otimes \sigma_j \varrho), \quad (i,j = x, y, z).\] 
In the above formula, we treat our qubits formally as spin-\(1/2\) particles. 
For the state \(\rho_{path}^{swap}\), \(T_{xx} = V_n\), \(T_{yy} = -V_n\), \(T_{zz}= 1\) and the rest are vanishing. 
A necessary and sufficient condition for a state of two qubits to violate local realism, for 
two settings per site, in the plane of \(\vec{n}\) and \(\vec{n}^{'}\)  is given by \cite{ZB}
\[
\sum_{i,j=\vec{n},\vec{n}^{'}}T_{ij}^2 >1.
\]
In our case (i.e. for the state \(\rho^{swap}_{path}\)), \(\sum_{i,j=x,y}T_{ij}^2 = 2V_n^2\), so that 
for 
\(V_n > \frac{1}{\sqrt{2}}\), one obtains a violation of local realism in the \(x-y\) plane. 
In the case of qubits defined by the output paths of the multiple entanglement
swapping devices considered here, \(x-y\) plane spin observables are equivalent 
to the measurement in the output ports of the external interferometers. 
That is, the  \(\vec{n} \cdot \vec{\sigma}\) operator for an \(\vec{n}\) in the \(x-y\) plane,
is equivalent to a device consisting of a phase shifter in front of 
a 50-50 beamsplitter and two detectors behind it.

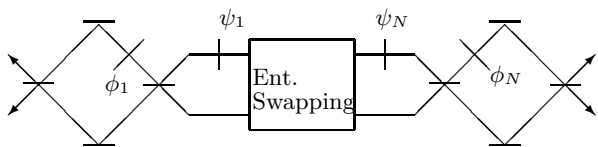
\begin{figure}[ht]
\begin{center}
\unitlength=0.4mm
\begin{picture}(-20,60)(0,0)
\put(-30, 10) {\line(1,0){35}}
\put(-30, -20){\line(0,1){30}}
\put(-30, -20){\line(1,0){35}}
\put(5, -20){\line(0,1){30}}

\put(-29, -5){Ent.} 

\put(-29, -13){Swapping}
 
\put(5, 5){\line(1,0){20}}
\put(15, 0){\line(0,1){10}}

\put(-40, 0){\line(0,1){10}}
\put(-40, 15){\(\psi_1\)}
\put(12, 15){\(\psi_N\)}
\put(-30, 5){\line(-1,0){20}}

\put(5, -15){\line(1,0){20}}
\put(-30, -15){\line(-1,0){20}}

\put(25,5){\line(1,-1){10}}
\put(25,-15){\line(1,1){10}}

\put(-50,-15){\line(-1,1){10}}
\put(-50,5){\line(-1,-1){10}}

\put(-65,-5){\line(1,0){10}}

\put(-75,0){\line(1,1){10}}

\put(-78,-7){\(\phi_{1}\)}
\put(-100,-5){\vector(-1,-1){10}}
\put(-100,-5){\vector(-1,1){10}}
\put(-105,-4.5){\line(1,0){10}}
\put(-100,-5){\line(1,1){20}}
\put(-100,-5){\line(1,-1){20}}
\put(-85,16){\line(1,0){10}}
\put(-85,-25){\line(1,0){10}}
\put(-60,-5){\line(-1,-1){20}}
\put(-60,-5){\line(-1,1){20}}
\put(50,0){\line(-1,1){10}}
\put(50, -4){\(\phi_N\)}

\put(75,-5){\vector(1,-1){10}}
\put(75,-5){\vector(1,1){10}}
\put(30,-4.5){\line(1,0){10}}
\put(35,-5){\line(1,1){20}}
\put(35,-5){\line(1,-1){20}}
\put(50,16){\line(1,0){10}}
\put(50,-25){\line(1,0){10}}
\put(75,-5){\line(-1,-1){20}}
\put(75,-5){\line(-1,1){20}}

\put(70,-5){\line(1,0){10}}

\end{picture}
\end{center}
\vspace{1.8cm}
\caption{The schematic diagram shows that to obtain better violation of local realism, the 
output ports of the entanglement swapping 
(contained in the square box in the figure) should be put to the Mach-Zehnder interferometers. } 
\label{figBV}
\end{figure}

The \(x-y\) plane does not  provide a violation of local realism for 
\(0\leq V_n \leq \frac{1}{\sqrt{2}}\), although the state is still entangled (and distillable) 
in that region. 
For these lower values of the visibility \(V_n\), 
we have to consider other planes for obtaining a violation. 
For example in the \(x-z\) plane, \(\sum_{i,j=x,z}T_{ij}^2 = 1+ V_n^2\), for the state 
\(\rho^{swap}_{path}\),
and thus a violation is always obtained in this plane. 
Hence, for these lower visibilities,
one must use other measurement planes to obtain a violation. Such violation can therefore be
obtained, only  by the  Mach-Zehnder interferometers at the output ports of 
the entanglement swapping device 
(Fig. \ref{figBV}). This is due to the fact that such a device is capable 
of performing any U(2) unitary transformation.

For a wide range of pumps and filters used in the swapping process, one will have 
\[ b=c.\]
For example, this is the case when the pulse and filter functions are Gaussian, i.e. 
\begin{equation}
\label{pump}
F(\omega_p) = 
\exp\{- \frac{(\omega_p -\omega_0)^2}{2 \sigma}\}
\end{equation}
 and 
\begin{equation}
\label{filter}
f(\omega_p) = 
\exp\{- \frac{(\omega_p -\omega_0/2)^2}{2 \sigma_f}\}.
\end{equation}

In such cases, the normalized state created after the idlers 
have been detected in the multiple entanglement swapping 
process with \(n\) PDCs is 
\[
\begin{array}{rcl}
\rho^{swap}_{path}&=& \frac{1}{2}(\left|00\right\rangle\left\langle 00\right| +
\left|11\right\rangle\left\langle 11\right| \\
&+& 
V_n^{(b=c)} \left|00\right\rangle\left\langle 11\right| 
+  V_n^{(b=c)} \left| 11 \right\rangle\left\langle 00\right|),
\end{array}
\]  
which can be rewritten as 
\[
\rho^{swap}_{path} = \frac{1 + V_n^{(b=c)}}{2}\left|\phi^{+}\right\rangle \left\langle \phi^{+}\right|
+ \frac{1 - V_n^{(b=c)}}{2}\left|\phi^{-}\right\rangle \left\langle \phi^{-}\right|
\]
where \(\left|\phi^{\pm}\right\rangle\) are the Bell states given by eq. (\ref{phipm}).

One has to bear in mind that the above results were obtained under the asuumption that we
deal with perfect entanglement swappings, especially no noise was allowed. 
It is obvious that for sufficiently 
low \(V_{n}\) (see Fig. \ref{figur}), both entanglement and 
violation of Bell inequalities would disappear, even for very minor noise admixtures.

\begin{acknowledgments}
A.S. and U.S. thank William J. Munro for helpful comments and 
are supported by  the University of Gda\'{n}sk, 
Grant No. BW/5400-5-0256-3. M.Z. acknowledges the KBN grant PBZ KBN 043/P03/2001.

\end{acknowledgments}

\appendix
\section{The two-photon state produced by  PDC}

This appendix is essentially meant to provide a ``derivation'' of the state (displayed in eqs. 
(\ref{pdcstate1}) and (\ref{pdcstate2})) produced in a parametric down conversion. It is
a reformulation of the theory of parametric down conversion which can be found in, e.g.,
\cite{Hong, Mandelbook}.

The phenomenon of spontaneous parametric down conversion (PDC) 
is a spontaneous fission of quasi monochromatic laser photons into  
correlated pairs of lower energy. All that takes place within a non-linear crystal. The probability 
of a single laser photon to fission is very low, 
but in a strong laser beam, the frequency of the phenomenon is quite high.  
The new photons, customarily called ``signal'' and ``idler'' have some basic properties.
First of all,
the wave vector ${\vec{k}}_{0}$ of the laser photons  is related to those (${\vec{k}}_{i}$ and ${\vec{k}}_{s}$)
of the idler and the signal by 
\({\vec{k}}_{0}\approx{\vec{k}}_{i}+{\vec{k}}_{s}\).
One has to stress that this relation holds within the crystal 
(and can be thought of as an approximate conservation of the linear momentum). 
Secondly, the frequencies $\omega_0$, $\omega_i$, and $\omega_s$ of the laser photon, idler and signal  satisfy  
\(\omega_{o}\approx\omega_{i}+\omega_{s}\).
And finally, the emerging pairs of photons are highly time-correlated. That is, if their
optical paths from the source to the detectors are equal (which we assume in this paper), the detection 
times are equal too (up to the time resolution of the detection system).
The relations between the wave vectors and between the frequencies 
are called phase matching conditions.

\subsection{Crystal-field interaction.}
Let us recall that in the interaction Hamiltonian
of the electromagnetic field with an atom or molecule, the 
dominating part is 
\(\hat{H}_{a-f}\sim\hat{\vec{\mu}}_{e}\cdot\vec{E}(\vec{x},t)\).
That is, it is proportional to the scalar product of the dipole moment of 
the atom or molecule with the local electromagnetic field.
Now the electric polarization of a material medium is given by the mean 
dipole moment of the atoms or molecules from which the medium is built 
(per unit volume).
 Let
$\vec{p}(\vec{x},t)$  stand for the local polarization 
of the volume $\delta V$, which contains the point $\vec{x}$ (the volume $\delta V$ 
is
macroscopically small but microscopically large).
The principal term of the interaction Hamiltonian for crystal and field
must have the form
\begin{equation}
H_{int}\sim\int_{V}\vec{p}(\vec{x},t)\cdot\vec{E}(\vec{x},t)d^{3}x,
\label{Hnl1}
\end{equation}
where $V$ is the volume of the crystal.
From the microscopic standpoint, the above formula reads
\begin{equation}
H_{int}\sim\int_{V}\sum_{n}\vec{\mu}_{n}\delta^{(3)}
(\vec{x}-\vec{x}_{n})\cdot\vec{E}(\vec{x},t)d^{3}x,
\label{hn1234}
\end{equation}
where $\vec{x}_{n}$ is a symbolic representation
of the position of the $n$-th  atom endowed with a dipole moment 
$\vec{\mu}_{n}$. 
The summation is over all atoms (the word ``atom'' standing for any stable
aggregate of charged particles, like atoms themselves, or ions, or molecules) of 
the medium. One can see that the formula
 (\ref{hn1234})  agrees with \(\hat{H}_{a-f}\).
If one introduces
the averaged (macroscopic) polarization
 (averaged over the macroscopically small volumes $\delta V$),
we get (\ref{Hnl1}).
One can assume that $\vec{E}(\vec{x},t)$ interacts with $\vec{p}(\vec{x},t)$
only in the point $\vec{x}$, thus the $i$-th
component of polarization is in the most general case given by
\begin{equation}
\begin{array}{rcl}
p_{i}(\vec{x},t)& =& \sum_{j=1}^{3}\chi_{ij}^{(1)}(\vec{x})E_{j}(\vec{x},t) \\
&+&
\sum_{j,k=1}^{3}\chi_{ijk}^{(2)}(\vec{x})E_{j}(\vec{x},t)E_{k}(\vec{x},t)+
\cdots,
\end{array}
\end{equation}
where $\chi_{ij}^{(1)}$ are  $\chi_{ijk}^{(2)}$ are the (macroscopic) polarizability 
coefficients. They are in the form of tensors. This is due to the 
fact that polarizablility may depend on the polarization
of the incoming light. 
Note here that for any crystal which is built
of molecules which are centro-symmetric the quadratic 
term of the polarizability vanishes.
Therefore the effect exists only in
birefringent media.

In the case of a perfect ``nonlinear'' crystal, we assume that
 $\chi_{ijk}^{(2)}(\vec{x})$ has the same value for all point 
within the crystal. 
The nonlinear term of the polarization gives the following
term in the interaction Hamiltonian (cf. 
(\ref{Hnl1})):
\[
\begin{array}{lcl}
&& H_{int}\sim\int_{V}\vec{p}(\vec{x},t)\cdot\vec{E}(\vec{x},t)d^{3}x \\
&& =  \int_{V}
\vec{p}^{\:lin}(\vec{x  },t)\cdot\vec{E}(\vec{x},t)d^{3} x \\
&& +  \int_{V}\vec{p}^{\:non}
(\vec{x},t)\cdot\vec{E}(\vec{x},t)d^{3}x,
\end{array}
\]
where $\vec{p}^{\:lin}$ ($\vec{p}^{\:non}$) 
is the linear (nonlinear) term of polarization.
The nonlinear part of the Hamiltonian is
\begin{equation}
H^{NL}\sim\int_{V}\sum_{ijk}\chi_{ijk}^{(2)}E_{i}(\vec{x},t)E_{j}(\vec{x},t)
E_{k}(\vec{x},t)d^{3}x. \label{Hnl2}
\end{equation} 
The quantized field can be 
expressed  (in the interaction picture) as
\begin{equation}
\begin{array}{lcl}
&& \vec{E}(\vec{x},t)=\sum_{p=1}^{2}\int\:d^{3}k\frac{i}{\sqrt{
2\omega(2\pi)^{3}}} \times \\
&& \hat{\epsilon}(\vec{k},p)a^-(\vec{k},p)\exp\{i(\vec{k}\cdot\vec{x}
-\omega t)\}+h.c.\\
&& =
\vec{E}^{(+)}(\vec{x},t)+\vec{E}^{(-)}(\vec{x},t),
\end{array}
\label{poleE}
\end{equation}
where
\(\vec{E}^{(-)}(\vec{x},t)=[\vec{E}^{(+)}(\vec{x},t)]^{\dagger}\),
and the summation is over two orthogonal linear polarizations, $h.c$ 
denotes the hermitian conjugate of the previous term, 
$\hat{\epsilon}(\vec{k},p)$ is a unit vector defining the 
linear polarization.
The symbol $a^-(\vec{k},p)$ 
stands for the annihilation operator of a photon of a wave vector
 $\vec{k}$ and polarization
$\hat{\epsilon}(\vec{k},p)$. The principal commutation rule for 
the creation and annihilation operators is given by
\([a(\vec{k},p),a^{\dagger}(\vec{k}',p')]=\delta_{p,p'}\delta^{(3)}(\vec{k}-\vec{k}')\), 
\([a^{\dagger}(\vec{k},p),a^{\dagger}(\vec{k}',p')]=0\) and 
\([a(\vec{k},p),a(\vec{k}',p')]=0\) .
As we are interested only in the PDC process,  we will neglect 
the depletion of the laser field and  assume that the total field
within the crystal is  
\(\vec{E}(\vec{x},t)=\vec{E}^{Laser}(\vec{x},t)+\vec{E}_{q}(\vec{x},t)\),
where the laser beam is described by a classical electromagnetic field
$\vec{E}^{Laser}$. In reality, the laser field is a mixture of coherent 
states, and its phase is undefined, but this is of no 
consequence to us here.
 The field $\vec{E}_{q}$ is described in a quantum-electrodynamical way.
It describes the secondary photons emitted by the crystal.
The down conversion takes place, thanks to the terms in
 (\ref{Hnl2}) of the  form
\(\int_{V}\sum_{ijk}\chi_{ijk}^{(2)}E_{i}^{Laser}E_{j}E_{k}d^{3}x\).
Only those terms of $E_{j}$ and  $E_{k}$
which contain the creation operators, can give rise to a two photon state, after acting on the vacuum state 
 $|\Omega\rangle$. The creation operators 
can be found only in the so-called negative frequency terms of the 
electromagnetic field operators (that is, in those which contain the 
factors
$\exp\{-i(\vec{k}\cdot\vec{x}-\omega t)\}$ (cf. (\ref{poleE}))).
Let us therefore forget about 
all other terms and analyze only 
\begin{equation}
H^{NL}\sim\int_{V}\sum_{ijk}\chi_{ijk}^{(2)}E_{i}^{Laser}E_{j}^{(-)}
E_{k}^{(-)}d^{3}x + \mbox{h.c.}
\label{Hnl4}
\end{equation}
For simplicity let us describe the laser field as a monochromatic wave
\(\vec{E}_{Laser}(\vec{x},t)=\hat{x}E_{0}(\exp\{i(\vec{k_0}\cdot\vec{x}-\omega_0
t-\phi)\}+c.c)\),
where $2E_{0}$ is the field amplitude and $c.c$ denotes
the complex conjugate of the previous expression. 
(Since an arbitrary electromagnetic field is a superposition of the
plane waves, it is very easy to get the general description.)
Then from 
(\ref{Hnl4}), one gets 
\begin{equation}
\begin{array}{lcl}
&& H^{NL} \sim \int_{V}\sum_{jk}\{\chi_{3jk}^{(2)}E_{0}\times \\
&&[\exp\{i(\vec{k}_{0}\cdot
\vec{x}-\omega_0 t-\phi)\}+c.c] \times \\
&& \sum_{p}\int d^{3}k\: f(\omega)\hat{\epsilon}(\vec{k},p)a^{\dagger}
(\vec{k},p)\exp\{-i\vec{k}\cdot\vec{x}\}\times \\
&& \sum_{p'}\int d^{3}k'\: f(\omega)\hat{\epsilon}(\vec{k},p)a^{\dagger}
(\vec{k}',p)\exp\{-i\vec{k'}\cdot\vec{x}\} d^{3}x \\
&& +  \mbox{h.c.},
\end{array}
\label{Hnl5}
\end{equation}
with $f(\omega)$ being a factor dependent on \(\omega\). Its specific structure is irrelevent here. 
Extracting only those elements of the above expression
which contain $\vec{x}$, one sees that their overall contribution
 to 
(\ref{Hnl5}) is given by 
\(\int_{V}d^{3}x\:\exp\{i\vec{x}(\pm\vec{k}_{0}-\vec{k}-\vec{k}')\}\), which we write as 
\(\Delta(\vec{k}_{0}-\vec{k}-\vec{k}')\).
If we assume that our crystal is a cube $L\times L\times L$, then for 
$L\rightarrow\infty$, \(\Delta\) 
approaches the Dirac delta \(\delta(\pm\vec{k}_{0}-\vec{k}-\vec{k}')\).
Thus, we immediately conclude that the emission
of the pairs of photons is possible only for the directions 
for which the condition 
\(\vec{k}_{0}\approx\vec{k}+\vec{k}'\) is met \cite{footnote1}.
Eq. (\ref{Hnl5}) can  be therefore put in the  form
\begin{equation}
\begin{array}{lcl}
&& H^{NL}\sim\sum_{p,p'}\int d^{3}k\int d^{3}k\Delta(\vec{k}_{0}-\vec{k}-\vec{k}')
A^{eff}_{p,p'} \times \\
&& \exp\{-i\omega_{0}t\}a^{\dagger}(\vec{k},p)a^{\dagger}
(\vec{k}',p') + \mbox{h.c},
\end{array}
\end{equation}
where \(A^{eff}_{p,p'} = \sum_{j,k}E_{0}\chi_{3jk}^{(2)}\hat{\epsilon}_{j}(\vec{k},p)\hat{\epsilon}_{k}
(\vec{k}',p')\)
is the effective
amplitude  of the laser pump field  
(which serves as a laser-crystal coupling strength).
This Hamiltonian fully describes the basic traits of the 
phenomenon of down conversion. In the so called type II down conversion, the laser pump beam is 
ordinary wheras the idler and signal photons are extraordinary. Thus we shall replace the 
general \(A^{eff}_{p,p^{'}}\) by \(F_0\).

\subsection{The state of photons emitted in the PDC process.}
We are interested in the process of production of pairs of photons.
Therefore we shall assume that the pump field is rather weak, so 
that the events of double pairs emissions are very rare. Therefore one
can use the perturbation theory.

The evolution of the state \(\left|\Psi_D(t)\right\rangle\) (in the interaction (Dirac) picture) of the photons 
emitted in the PDC process is governed by 
 the Schr\"odinger equation 
\(i\hbar\frac{d}{dt}|\Psi_{D}(t)\rangle=H^{NL}_{D}(t)|\Psi_{D}(t)\rangle\),
where
\(H^{NL}_{D}(t)=\exp\{\frac{i}{\hbar}H_{0}t\}H^{NL}\exp\{-\frac{i}{\hbar}H_{0}t\}\),
and \(H_0 + H^{NL}\) is the total Hamiltonian of the system.
Therefore
\begin{equation}
\begin{array}{rcl}
|\Psi_{D}(t)\rangle&=& |\Psi_{D}(t_{0})\rangle + \frac{1}{i\hbar}
\int_{t_{0}}^{t}H^{NL}_{D}(t')|\Psi_{D}(t')\rangle dt' \\
& \simeq & |\Psi_{D}(t_{0})\rangle
+\frac{1}{i\hbar}
\int_{t_{0}}^{t}H^{NL}_{D}(t')|\Psi_{D}(t_{0})\rangle dt',
\end{array}
\label{disturb}
\end{equation}
where we have replaced the time 
dependent state in the integral by its initial form 
$|\Psi_{D}(t_{0})\rangle$ using the first order of the perturbation calculus.
In the Dirac picture, the creation and annihilation operators
depend on time as
\(a_{D}^{\dagger}(\vec{k},p,t)=\exp\{i\omega t\}a^{\dagger}(\vec{k},p)\)
and \(a_{D}(\vec{k},p,t)=\exp\{i\omega t\}a(\vec{k},p)\).
In (\ref{disturb}), we put  $t_{0}=0$, and we take
the vacuum state (no photons) $|\Omega\rangle$ as the initial state \(|\Psi(0)\rangle\).
We are interested only in the term with the integral, because it is 
only there that one can find creation operators responsible for 
the spontaneous emission of pairs of photons.
The photons interact with the laser field only during the time of the order of
 $\frac{L}{c}$.  The interaction simply ceases when they leave the crystal.
Therefore, as the annihilation operators 
when acting on the vacuum state give zero, one can write
\[
\begin{array}{lcl}
&& \sum_{p,p'}\int d^{3}k\int d^{3}k^{'} F_{0} \Delta(\vec{k}_{0}-\vec{k}-\vec{k}')\Delta
(\omega+\omega'-\omega_{0})\\
&& a^{\dagger}(\vec{k},p)a^{\dagger}(\vec{k}',p')
|\Omega\rangle,
\end{array}
\]
where we have written \(\int_{0}^{\frac{L}{c}}dt'\exp\{it'(\omega+\omega'-\omega_{0})\}\)
as \(\Delta
(\omega+\omega'-\omega_{0})\). As \(L \rightarrow \infty\), \(\Delta\) behaves as a Dirac delta.
Thus the allowed frequencies of the emissions satisfy the 
relation \cite{footnote2}
$\omega_{0}\approx\omega+\omega'$.
Since $\frac{L}{c}\omega$, $\frac{L}{c}\omega'$ and 
$\frac{L}{c}\omega_{0}$ are typically of the order of
$10^{4}$ the function $\Delta(\omega+\omega'-\omega_{0})$ 
is very close to
$\delta(\omega+\omega'-\omega_0)$.

\subsection{Directions of emissions}

We know that
\(\omega=|\vec{k}|\frac{c}{n(\omega,p)}\),
where $\frac{c}{n(\omega,p)}=c(\omega,p)$ is the speed of light 
in the given medium, which depends on frequency and the polarization.
 Using this relation as well as phase matching condition for frequencies,
we get the condition
\(
|\vec{k}_{0}|c(\omega_{0})\simeq|\vec{k}|c(\omega,p)+|\vec{k}'|c(\omega',p')\).
Therefore, the emissions of the pairs are possible only when 
the phase matching and the above condition are both met. 
This means that the correlated emissions occur only for specific 
directions, specific frequencies and specific polarizations.
There are two types of down conversions:
\begin{enumerate}
\item both photons of a pair have the same polarization ({\em type {\sc I}}),
\item they have orthogonal polarizations ({\em type {\sc II}}).
\end{enumerate}   
Additionally if one has:
\begin{enumerate}
\item $\omega\simeq\omega'$ 
then we have a frequency degenerate PDC,
\item and if $\hat{k}=\hat{k}'$, 
then we have a co-linear one.
\end{enumerate}

\subsection{Time correlations}

In this section it will be shown that the down converted photons 
are very tightly time correlated.
The probability
of a detection of a photon, of say, the horizontal polarization $H$,
at a detector situated at point  $\vec{x}$ and at the
moment of time $t$, is proportional to
\begin{equation}
p(\vec{x},t,H)\simeq\eta Tr\varrho E_{H}^{(-)}(\vec{x},t)E_{H}^{(+)}(\vec{x},t),
\label{pro10}
\end{equation}
where $\eta$ 
is the coefficient which characterizes the quantum efficiency of the 
detection process,
$\varrho$ is the density operator, which represents the field 
in the Heisenberg picture,  $E_{H}$ is the horizontal component of the field 
in the detector. For the above relation to be 
true, we also assume that via the aperture of the detector enter 
only photons of a specified direction of the wave vector.  

For a pure state, (\ref{pro10}) reduces to
\(p(\vec{x},t,H)\simeq\langle\psi|E_{H}^{(-)}E_{H}^{(+)}|\psi\rangle\).
The probability of a joint detection of two photons, of polarization
 $H$, at the locations $\vec{x}_{1}$ and  $\vec{x}_{2}$, and at the 
moments of time $t_1$ and $t_2$ is proportional to
\begin{equation}
\begin{array}{lcl}
&& p(\vec{x}_{1},t_1;\vec{x}_{2},t_2) \sim\\
&& \langle\psi|E_{H}^{(-)}(\vec{x}_{1},t_1)
E_{H}^{(-)}(\vec{x}_{2},t_2)E_{H}^{(+)}(\vec{x}_{2},t_2)E_{H}^{(+)}(\vec{x}_{1},t_1)
|\psi\rangle.
\label{pro11}
\end{array}
\end{equation}
The evolution of the field within the crystal lasts for the time around
$\frac{L}{c}$.
If the detectors are very far away from each other, and from the crystal, 
then the photon field can be treated as free-evolving.
State $|\psi\rangle$ is the photon state that leaves the crystal 
at the moment around
$\frac{L}{c}$:
\begin{equation}
|\psi(t=\frac{L}{c})\rangle=|\Omega\rangle+\frac{1}{i\hbar}e^{-\frac{i}{\hbar}H_{0}
\frac{L}{c}}\int_{0}^{\frac{L}{c}}H_{D}^{NL}(t')dt'|\Omega\rangle.
\end{equation}
Let $t=t_1$ i $t_2=t'$, and $|\psi\rangle=|\psi(t=\frac{L}{c})\rangle$, 
then 
(\ref{pro11}) can be written down as
\begin{equation}
\begin{array}{lcl}
&& p(\vec{x}_{1},t|\vec{x}_{2},t')\simeq \\ 
&&\langle\psi|E_{H}^{(-)}
(\vec{x}_{1},t)E_{H}^{(-)}(\vec{x}_{2},t')E_{H}^{(+)}(\vec{x}_{1},t)
E_{H}^{(+)}(\vec{x}_{2},t')|\psi\rangle.
\label{pro12}
\end{array}
\end{equation}
If we choose just two conjugate propagation directions (i.e. such 
that fulfill the phase-matching conditions), 
then approximately one has 
\[
\begin{array}{rcl}
E_{H}^{(+)}(\vec{x},t)&= &\int d\omega e^{-i\omega t}f_1(\omega)a_1(\omega)\\ 
E_{H}^{(+)}(\vec{x}',t')&= & \int d\omega e^{-i\omega t'}f_2(\omega)a_2(\omega),
\end{array}\]
where $f_{1}$ and $f_2$ are
the frequency response function 
which characterize
the detections (or rather filter-detector system). 
We assume that the response functions are such 
that their maxima agree with the frequencies determined by the 
phase-matching conditions.
The annihilation and creation operators which were used above,
are replaced by $a_{i}(\omega)$ and its conjugate, 
which can be used to describe ``unidirectional'' excitations of the 
photon field (i.e., we assume that the detectors see only 
the photons of a specified duration of propagation, this a good assumption
if the detectors are far from the crystal, and
the apertures are narrow). The index  $i$ defines the direction (fixed)
of the wave vector. The operators satisfy commutation relation, which are
a modification of those given above to the current specific case 
\([a_{i}(\omega),a^{\dagger}_{j}(\omega')]=\delta_{ij}\delta(\omega-\omega')\),
\([a_{i}(\omega),a_{j}(\omega')]=0\). 

Introducing  an unit operator
\(\hat{I}=\sum_{i=0}^{\infty}|b_i\rangle\langle b_i|\),
where $|b_i\rangle$ is a basis state into (\ref{pro12}), we obtain
\begin{equation}
\begin{array}{lcl}
&& p(\vec{x}_{1},t|\vec{x}_{2},t')\simeq\\ 
&&\langle\psi|E_{H}^{(-)}
(\vec{x}_{1},t)E_{H}^{(-)}(\vec{x}_{2},t')\hat{I}E_{H}^{(+)}(\vec{x}_{1},t)
E_{H}^{(+)}(\vec{x}_{2},t')|\psi\rangle.
\label{pro13}
\end{array}
\end{equation}
Since
 $E_{H}^{(+)}$  contains only 
the annihilation operators,
they transform the two photon state $|\Psi\rangle$ 
into the vacuum state. Thus, the equation 
(\ref{pro13}) can be written down as
\begin{equation}
p(\vec{x}_{1},t|\vec{x}_{2},t')\simeq\langle\psi|E_{H}^{(-)}
E_{H}^{'(-)}|\Omega\rangle\langle\Omega|
E_{H}^{(+)}E_{H}^{'(+)}|\psi\rangle,
\label{pro14}
\end{equation} 
where the primed expressions pertain to the moment of time $t'$ and 
the position
$\vec{x}_{2}$. Thus we have 
\(p(\vec{x}_{1},t|\vec{x}_{2},t')\simeq |A_{12}(t,t')|^{2}\), where 
\(A_{12}(t,t')=\langle\Omega|E_{H}^{(+)}(\vec{x}_{1},t)
E_{H}^{'(+)}(\vec{x}_{2},t')|
\psi\rangle\).
The state $|\Psi\rangle$ can be approximated by  
\begin{equation}
|\Psi\rangle=
|\Omega\rangle+\int d\omega_1\int
d\omega_{2}F_{0}\delta(\omega-\omega_1-\omega_2)a^{\dagger}_{1}
(\omega_1)a^{\dagger}_{2}(\omega_2)|\Omega\rangle.
\end{equation}
Then
\begin{equation}
\begin{array}{lcl}
&& A_{12}(t,t')=\langle\Omega|\int d\omega'e^{-i\omega't}f_{2}(\omega')a_{2}(\omega')\\ \nonumber
&& \int d\omega e^{-i\omega t}f_{1}(\omega)a_{1}(\omega) \int d\omega_2\int d\omega_2 F_{0}
\delta (\omega_0-\omega_1-\omega_2)\\ \nonumber
&& a_{2}^{\dagger}.
(\omega_2)a_{1}^{\dagger}(\omega_1)|\Omega\rangle
\end{array}
\end{equation}
Since the creation and annihilation operators 
for different modes commute, and since one can use  
\(\langle\Omega|a_{i}(\omega')a_{j}^{\dagger}(\omega)|\Omega\rangle=
\delta_{ij}\delta(\omega'-\omega)\),
we get
\begin{equation}
A_{12}(t,t')=e^{-i\omega_0 t'}\int d\omega F_{0}
e^{-i\omega(t-t')}f_2(\omega_0-\omega)f_1(\omega).
\end{equation}
If \(F_0\) varies very slowly, 
which is a  good asumption in the case of a crystal,
then we have 
\begin{equation}
\begin{array}{lcl}
&& p(\vec{x}_{1},t|\vec{x}_{2},t')\sim |A_{12}(t,t')|^{2}\\
&&\simeq|
\int d\omega e^{-i\omega(t-t')}f_2(\omega_0-\omega)f_1(\omega)|^{2}.
\label{pro16}
\end{array}
\end{equation}
We see that the probability will depend on the difference 
of the detection times.

To illustrate the above, let us assume that $f_{1}=f_{2}=f$, and that they are gaussian. 
Then, if one assumes that the central frequency of $f$ is 
$\omega_c=\frac{1}{2}\omega_0$ and
\(f(\omega)=Ce^{-\frac{(\omega_c-\omega)^{2}}{{\sigma}^{2}}}\),
then we have \(f_{1}(\omega)=f_{2}(\omega_0-\omega)=f(\omega)\).
If one uses these relations the probability
of detection of two photons at the moments
$t$ and $t'$ satisfies the following
dependence
\begin{equation}
\begin{array}{lcl}
&& p(\vec{x}_{1},t|\vec{x}_{2},t')\\
&& \sim |\int d\omega e^{-i\omega(t-t')}C^{2}e^
{-2\frac{{(\omega_c-\omega)}^{2}}{{\sigma}^{2}}}|^{2}\sim
e^{-\frac{\sigma^{2}}{2}(t-t')^{2}}.
\label{pro15}
\end{array}
\end{equation}
We see, that if $\sigma\rightarrow\infty$ (that 
is, the detector has an infinitely broad frequency response) 
then the expression (\ref{pro15}) approaches
the Dirac delta
$\delta (t-t')$ and this means, 
that the two detectors register the two photons at the same 
moment of time.
However, such detectors do not exist.
Nevertheless, from equations
(\ref{pro16}) and (\ref{pro15}), 
we see that the degree of time correlation of the detection 
of the PDC photons
depends on the frequency response of the detectors.
Thus, the photons are almost perfectly time correlated. 

We have shown what are the reasons for the 
properties of the PDC photons. Although the above reasoning was 
done under the assumption of a monochromatic nature of the pumping field,
all this can be generalized to the 
non-monochromatic case, including the most interesting one of a pulsed pump.
The distinguishing traits of this situation can be summarized
by the following remarks. The emission from the crystal can appear only
when the pulse is within the crystal. Further, the frequency
$\omega_0$ and the wave vector are
not strictly defined. If the pulse is too short
because of the relation $T\approx\frac{1}{\Delta\omega}$, where
$T$ is the pulse width, 
the PDC photons are less tightly correlated directionally.

The two photon state coming out of  a PDC can then be approximated by  
\begin{equation}
\label{gera}
\begin{array}{lcl}
|\Psi\rangle= && \int d\omega_0 F(\omega_0)
\int d\omega_1\int
d\omega_{2} \times \\
&& \Delta(\omega_0-\omega_1-\omega_2)a^{\dagger}_{1}
(\omega_1)a^{\dagger}_{2}(\omega_2)|\Omega\rangle,
\end{array}
\end{equation}
 where we have replaced the effective pump amplitude 
by  the spectral decomposition of the laser pulse \(F(\omega_0)\). Since a pulse is a 
superposition of monochromatic waves, we therefore integrate over the spectrum.


\begin{thebibliography}{99}

\bibitem{ZZHE}M. \.Zukowski, A. Zeilinger, M.A. Horne and A.K. Ekert, Phys. Rev. Lett.
\textbf{71}, 4287 (1993).
 
\bibitem{SougatoSwap}S. Bose, V. Vedral, and P.L. Knight, Phys. Rev. A \textbf{57}, 822 (1998);
\emph{ibid.} \textbf{60}, 194 (1999).

\bibitem{tele}  C.H. Bennett, G. Brassard, C. Crepeau, R. Josza, A Peres, and  W.K. Wootters,
Phys. Rev. Lett. \textbf{70}, 1895 (1993).

\bibitem{Pan} J.-W. Pan, D. Bowmeester, H. Weinfurter, and A. Zeilinger, Phys. Rev. Lett. \textbf{80}, 
3891 (1998).


\bibitem{ZukowskiSwapProc} M. \.Zukowski, A. Zeilinger, and H. Weinfurter,
Annals N.Y. Acad. Sci. \textbf{755}, 91 (1995).



\bibitem{Briegel}  H.-J. Briegel, W. D{\" u}r, J.I. Cirac, and P. Zoller, Phys. Rev. Lett. \textbf{81}, 5932 (1998);
 W. D{\" u}r, H.-J. Briegel, J.I. Cirac, and P. Zoller, Phys. Rev. A \textbf{59}, 169 (1999).

\bibitem{huge}  C.H. Bennett, D.P. DiVincenzo, J.A. Smolin, and W.K. Wootters, Phys. Rev. A \textbf{54}, 3824 (1996).

\bibitem{ZHWZ} A. Zeilinger, M.A. Horne, H. Weinfurter, and M. \.Zukowski, Phys. Rev. Lett.
\textbf{78}, 3031 (1997).


\bibitem{footnote_noi}We shall not 
discuss here the effects linked with the statistics of the output of the PDC sources.

\bibitem{PT}The partial
 transpose of a bipartite state \(\rho_{AB} = \rho_{i\mu j\nu}\), where
 \(i\) and \(j\) are the indices for party \(A\) and \(\mu\) and \(\nu\) are
 the indices for party \(B\), with respect to part \(A\) is \(\rho^{T_{A}}_{AB}
 = \rho_{j\mu i\nu}\) \cite{Peres1}. A state \(\rho_{AB}\) is
 said to have positive partial transpose  if \(\rho_{AB}^{T_A}\) is
 a positive operator. \(\rho_{AB}\) has non-positive partial transpose otherwise.
 A state with non-positive partial transpose is always entangled \cite{Peres1}.


 \bibitem{Peres1}A. Peres, Phys. Rev. Lett \textbf{77} 1413 (1996).
 
\bibitem{HHHPPT}M. Horodecki, P. Horodecki, and R. Horodecki, Phys. Lett. A \textbf{223}, 1 (1996).



\bibitem{Wootters} S. Hill and W.K. Wootters, Phys. Rev. Lett. \textbf{78}, 5022 (1997); W.K. Wootters, Phys. Rev. Lett.
\textbf{80}, 2245 (1998).

\bibitem{Cirac}  G. Vidal, W. D{\" u}r, and J.I. Cirac, Phys. Rev. Lett. \textbf{89}, 027901 (2002).

\bibitem{ratio} M. Horodecki, A. Sen(De), and U. Sen, \emph{
The rates of asymptotic entanglement transformations for bipartite mixed states: 
Maximally entangled states are not special}, to appear in Phys. Rev. A (quant-ph/0207031).

\bibitem{Shor}  P.W. Shor, J. Math. Phys. \textbf{43}, 4334 (2002).

\bibitem{HHHdist} M. Horodecki, P. Horodecki, and R. Horodecki, Phys. Rev. Lett. \textbf{78}, 574 (1997).


\bibitem{Rains1}E.M. Rains, 
Phys. Rev. A \textbf{60}, 179 (1999);
E.M. Rains, Phys. Rev. A \textbf{63}, 019902 (2001); E.M. Rains, \emph{A semidefinite 
program for distillable entanglement}, quant-ph/0008047.

\bibitem{Rains2}E.M. Rains, Phys. Rev. A \textbf{60}, 173 (1999) .



\bibitem{Bell}J.S. Bell, Physics \textbf{1}, 195 (1964).

\bibitem{HHHBV} R. Horodecki, P. Horodecki, and M. Horodecki, Phys. Lett. A \textbf{200}, 340 (1995).

\bibitem{ZB} M. \.Zukowski and {\v C}. Brukner, Phys. Rev. Lett. \textbf{88}, 210401 (2002).

\bibitem{Hong}C.K. Hong and L. Mandel, Phys. Rev. A \textbf{31}, 2409 (1985).

\bibitem{Mandelbook}L. Mandel and E. Wolf, \emph{Optical Coherence and Quantum Optics}, 
Cambridge University Press (1995). 

\bibitem{footnote1} The 
condition with the minus sign in the delta function 
cannot be met because in this case the phase matching condition for 
the frequancies cannot be satisfied.

\bibitem{footnote2} Note here that if we had  kept in the Hamiltonian 
the 
terms for which 
\(
-\vec{k}_{0}\approx\vec{k}+\vec{k}'
\), then the  condition $-\omega_{0}\approx\omega+\omega'$ for frequencies would have emerged (cf. Ref. \cite{footnote1}), 
and this is  \emph{impossible} to meet! 

\end{thebibliography}
\end{document}